# Properties of Individual Dopant Atoms in Single-Layer MoS$_2$: Atomic Structure, Migration, and Enhanced Reactivity

Yung-Chang Lin,[1] Dumitru O. Dumcenco,[2] Hannu-Pekka Komsa,[3] Yoshiko Niimi,[1] Arkady V. Krasheninnikov,[3,4] Ying-Sheng Huang,[2] Kazu Suenaga[1]*

[1]National Institute of Advanced Industrial Science and Technology (AIST), Tsukuba 305-8565, Japan

[2]Department of Electronic Engineering, National Taiwan University of Science and Technology, Taipei 10607, Taiwan

[3]Department of Physics, University of Helsinki, P.O. Box 43, Fl-00014 Helsinki, Finland

[4]COMP/Department of Applied Physics, Aalto University, P.O. Box 11100, FI-00076 Aalto, Finland

**ABSTRACT**

The differences in the behavior of Re (n-type) and Au (p-type) dopant atoms in single-layered MoS$_2$ were investigated by *in situ* scanning transmission electron microscopy. Re atoms tend to occupy Mo sites, while Au atoms exist as adatoms and show larger mobility under the electron beam. Re substituted to Mo site showed enhanced chemical affinity, evidenced by agglomeration of Re adatoms around these sites. This may explain the difficulties in achieving a high compositional rate of homogeneous Re doping in MoS$_2$. In addition, an *in situ* coverage experiment together with density functional theory calculations discovered a high surface reactivity and agglomeration of other impurity atoms such as carbon at the Re doped sites.

Dopants generally govern physical and chemical properties of all kinds of solids, especially semiconductors where doping is widely used to adjust carrier densities and tailor the electronic characteristics of the devices. The effects of doping become more significant in case of two-dimensional (2D) materials, as in addition to the substitutional doping, carrier concentration can be controlled by adding adatoms or layers of another material. For the case of graphene, the Fermi level can be tuned efficiently by metal contact[1] or noncovalent chemical functionalization.[2] Recently, layered inorganic transition-metal dichalcogenides (TMDs) exhibiting a wide variety of electronic properties[3-5] have received lots of attention. Among them, single-layer $MoS_2$ possesses a direct gap[6,7] and exhibits carrier mobility comparable to the graphene nano-ribbons with high current on-off ratios.[8] The electrical conductivity of $MoS_2$ may then be further modulated by substitutional doping, such as Re (n-type)[9-11] and Nb (p-type).[12,13] The $MoS_2$ transistors doped with Au show a remarkable p-type character.[14] Besides, intercalation of noble metals (Au or Ag) in $MoS_2$ nanotubes has been reported[15,16] so that the knowledge of how dopant atoms interact with the host TMD crystal lattices is highly important in the context of both fundamental science and future device technology.

The Z-contrast annular dark-field (ADF) imaging by means of STEM is a powerful method to identify not only the structure but also chemical composition atom-by-atom. Voyles[17] visualized individual Sb atoms and clusters in bulk Si, and our previous work reported a random distribution of hetero-transition metals in the single-layer $Mo_{1-x}W_xS_2$ alloy.[18] Recently, single three or four coordinated Si impurity atoms in graphene lattice were also reported.[19,20] Finally, the presence of Co and Ni dopants at the edges of $MoS_2$ hydrotreating catalysts was demonstrated in scanning tunneling microscopy studies.[21,22] However, no direct experimental information on the atomic configurations of dopants in 2D TMDs is available so far. Moreover, correlations between their dynamic behavior and nano-scale properties have never been studied before.

Here we perform an STEM study focused on the atomic structures of single-layered $MoS_2$ doped with Re and Au. By combining the experimental STEM observations with first-principles calculations carried out within the framework of the density functional theory (DFT), we unequivocally identify the atomic positions of Re and Au dopants in the host $MoS_2$ lattice and correlate the migration behaviors of dopant atoms with their energetics. Dynamic observations of the atomic structure evolution reveal the interaction of the dopant atoms with impurity atoms deposited on the doped $MoS_2$ sheet, which proves local enhancement of the chemical affinity of the system.

Typical ADF images of single-layered Re-doped and Au-doped $MoS_2$ are presented in Fig. 1a and 1b, respectively. The dopants, Re ($Z = 75$) and Au ($Z = 79$), appear in brighter contrast in the ADF images than both Mo ($Z = 42$) and S ($Z = 16$) as indicated by arrows. Chemical analysis by means of EDX was also done to confirm the doping elements (Fig. S2 and S3). The high resolution ADF image in the inset of Fig. 1a clearly shows that Re atoms sit at the Mo sites. The Re dopants are well dispersed in $MoS_2$ layers and seldom form clusters on the host material. The observed doping concentration is around ~0.6 at% Re in agreement with the synthesis condition. On the other hand, the Au dopants at similar concentration (~0.6 at% Au) tend to aggregate on the $MoS_2$ surface (open circles). The Au atoms are mobile under the electron beam and can be found either on top of the Mo, S, or the hollow center (HC) of $MoS_2$ hexagons, as shown in the inset of Fig. 1b.

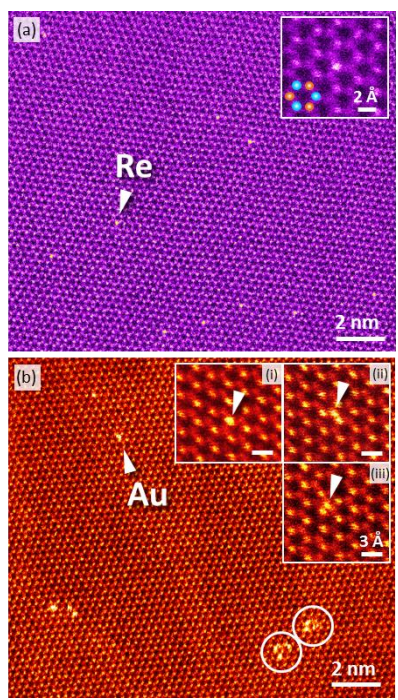

**Figure 1.** (a) Large area ADF image of clean Re-doped single-layer $MoS_2$. The inset shows a Re dopant substitution at the Mo site with atomic resolution as evident from a brighter contrast. Blue and orange balls indicate Mo and two overlapped S atoms, respectively. There are 10 Re atoms (one pointed by the arrow) found in ~1500 $MoS_2$ unit cells, which corresponds to a doping concentration of ~0.6 at%. (b) Large area ADF image of ~0.6 at% Au-doped single-layer $MoS_2$. The insets (i), (ii), and (iii) show Au adatom (indicated by arrows) located on top of the Mo, S, and HC site, respectively.

In order to understand the dynamic behaviors of Re dopants in $MoS_2$, we analyzed ~450 Re dopant atoms. 89% of the Re atoms (~400) are substitutional dopants at the Mo sites (Re@Mo), and they did not migrate at all throughout the acquisition time (typically 240 s). We found that the rest 11% of Re were adatoms (~51) and changed their positions in sequential images. Figs. 2(a)-2(c) show a set of sequential ADF images illustrating the migration of four Re dopant atoms in a $MoS_2$ single-layer. Three Re atoms (Re@Mo) remained stationary for the whole acquisition time (up to t = 335.4 s). In Fig. 2(b), the Re adatom on S (Re-S, indicated by open arrow) jump to the Mo site (Re-Mo) and then move to another S site (Re-S) (Fig. 2(c)) to form again the complex Re-S+Re@Mo dopant structure (Movie M1). Interestingly, the Re-S structure always appears in the vicinity of Re@Mo, and therefore forms the dopant structure as a complex of Re-S+Re@Mo among 17 cases observed. The experimental ADF image (Fig. 2(h)) was corroborated by STEM simulation with DFT calculated atomic model (Fig. 2(g)) using the appropriate the experimental parameters. The simulated ADF profile shows a perfect match with the experiment (Fig. 2(i)), if possible specimen inclination is taken into account. Other observed Re doped structures are summarized in Fig. S4.

On the contrary, 95% of Au dopants are highly mobile adatoms easily migrating under the electron beam, and the rest 5% of Au dopants are stable Au@Mo complexes in the dataset of 220 Au dopant atoms observed. Figs. 2(d)-2(f) show sequential ADF images of Au migrations (Movie M3). One Au-S atom in the single layer region is continuously moving and has a different position in each frame. Another Au atom continues to migrate in the bilayer region. The bilayer $MoS_2$ region shows unconventional rhombohedral stacking (Fig. S6), which probably comes from the intercalation of Au atoms.[23] The migration behavior of the Au atom can be interpreted as it is migrating between the layers.[15,16] Figs. 2(j)-2(l) show the identification of Au-S structure by comparing the ADF profile with the STEM simulation. The other Au dopant structures are summarized in Fig. S5. Statistically, most of the Re atoms are in stable Re@Mo positions, and Au atoms are mobile adatoms (Fig. S7).

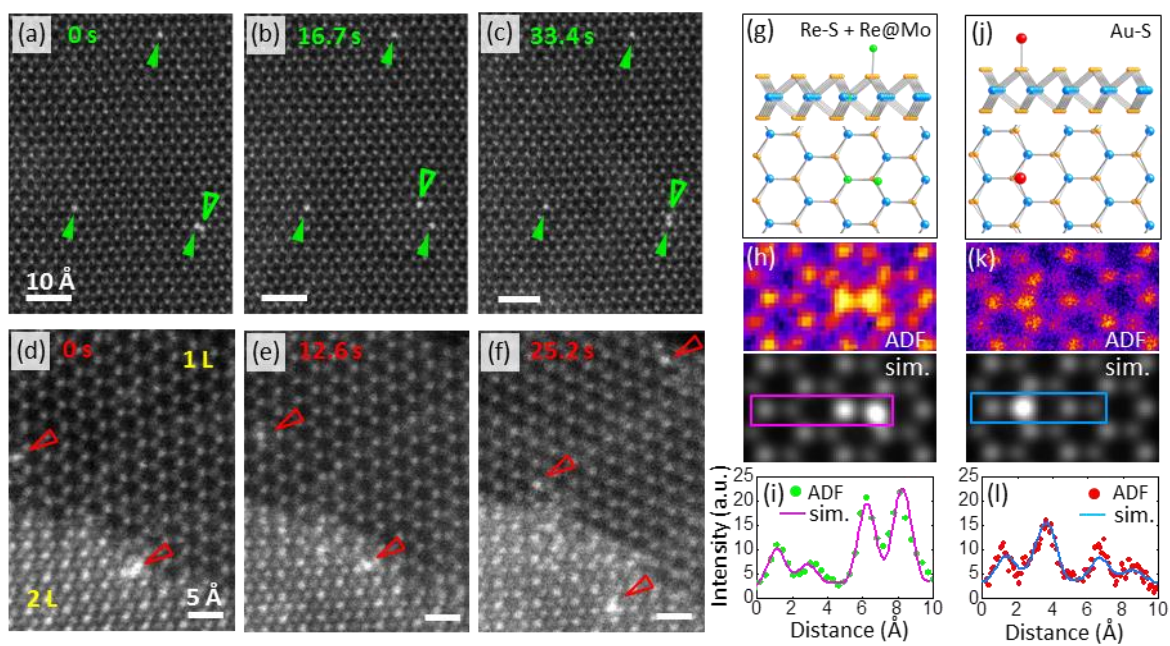

**Figure 2.** Sequential ADF images of single-layer Re-doped and Au-doped MoS$_2$. (a) Three Re@Mo are indicated by green arrows and one Re-S adatom is indicated by the open arrow. (b) At t = 16.7 s, the Re-S adatom moves to the Mo site as a Re-Mo and (c) move to another S site as a Re-S again at t = 28.5 s. (d-f) Au adatoms on single- and bilayer regions (indicated by red open arrows) change their atomic positions in every frame. The bilayer region shows rhombohedral stacking [see Fig. S4]. (g)(j)The atomic model of Re-S+Re@Mo complex dopant structure and Au-HC. (h)(k) The corresponding ADF images and simulated images of Re-S+Re@Mo and Au-HC. (i)(l) The profile of experimental ADF counts and fitting of simulation line profiles for the two cases.

To confirm the migration behavior of the dopant atoms under electron beam, we carried out spin-polarized DFT calculations of doped MoS$_2$ layers using the VASP simulation package[24,25] as described in supporting information and computed the energetics of dopant atoms in the subsitutional and adatom positions. The formation energies ($E_f$) of a dopant atom in adatom and substitutional configurations was calculated as

$$E_f = E_{total}^D - (E_{total} - \delta\mu_Y + \mu_D) \qquad (1)$$

where $E_{total}^D$ and $E_{total}$ are the total energies of the doped and pristine supercells; $\mu_D$ and $\mu_Y$ are the chemical potentials of a single dopant (D = Re or Au) and the substituted atom in the host lattice (Y = Mo or S), respectively. $\mu_Y$ was chosen to correspond to the bulk phase of each element. In the case of adatom, $\mu_Y$ term was omitted: $\delta = 0$; while $\delta = 1$ in Eq. (1) corresponds to the case of the substitutional dopant.

**TABLE I.** Formation energy of Re and Au adatom/substitutional defect complexes on Mo, S, and HC sites, as calculated using DFT. $E_k$ is the maximum kinetic energy of each dopant atom received from 60 kV electron beam.

| Dopant | $E_k$ (eV) | Adatom (eV) | | | Sub. (eV) | |
|---|---|---|---|---|---|---|
| | | Mo | S | HC | Mo | S |
| Re | 0.75 | 6.09 | 7.37 | 7.28 | 1.9 | 6.0 |
| Au | 0.71 | 2.46 | 2.33 | 2.37 | 7.2 | 3.1 |

The formation energies of dopant atom in different configurations are presented in Table I. The values of $E_f$ give an idea of what the favored sites for the dopants are after high-temperature CVT growth, which is relatively close to thermal equilibrium conditions. In the case of Re dopants, substitution to Mo site has the lowest formation energy by a large margin. Among the adatom sites, the position over Mo is favored. This is in a good agreement with the observations that nearly all Re atoms are located at Mo sites, and rarely on the adatom sites on top of Mo and S. In a stark contrast, substituting

Mo with Au is highly unfavorable, and the Au atoms are expected to be located predominantly at the adatom sites, which is also consistent with our experimental findings. Several possible configurations for the complex structures with two Re atoms were considered (Table S1). The complex dopant structure is energetically more stable than the two isolated ones according to the DFT calculations.

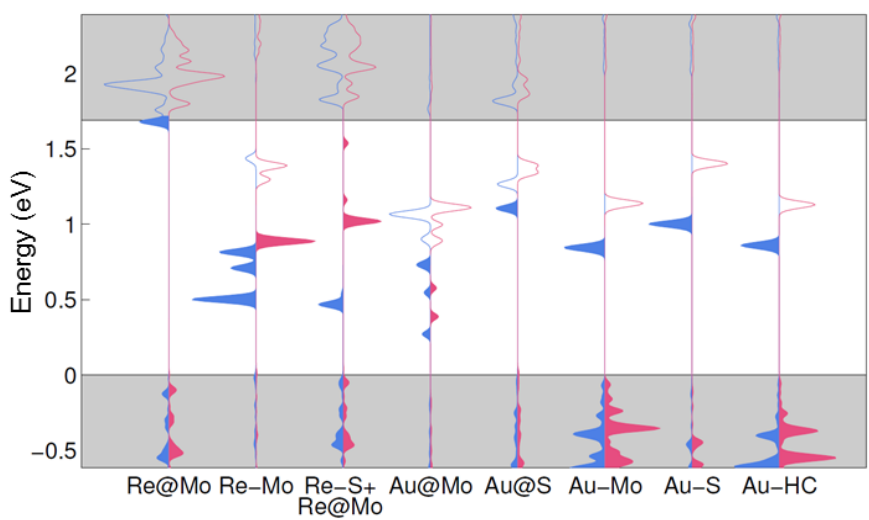

**Figure 3.** Local density of states of the Re and Au dopant atoms at eight considered atomic structures. The blue and red lines represent the spin-up and spin-down components, and the electron occupied states are shown as filled curves. The grey areas denote the valence band (lower) and conduction band (upper) energy ranges.

Local densities of states (LDOS) around the dopant atoms were also calculated for all the experimentally found atomic configurations including the very rare occasions as shown in Fig. 3. Re dopants substituting Mo atoms provide extra electrons into the conduction band of the $MoS_2$ (n-type). On the other hand, Au dopants in/on $MoS_2$ give rise to similar mid-gap states at all the sites and may compensate n-type doping. For larger Au clusters, the work function is expected to lie close to valence band, subsequently leading to p-type doping, as observed in Ref. 14. Very interestingly, our simulations suggest that Au adatoms can possess magnetic moments on $MoS_2$ surfaces, so that the introduction of Au impurities may lead to new functionalities of $MoS_2$ and likely other TMDs. Moreover, gold nanoparticles on $MoS_2$ were reported to dramatically enhance the photocurrent.[26]

In order to prove that the dopants indeed alter the local physical/chemical properties of $MoS_2$ layers, carbon deposition was introduced after 15 min of e-beam shower and the interaction between carbon contaminations and dopants was recorded afterwards in a reasonable time scale. In Fig. 4, a sequence of STEM images illustrates the *in situ* observation of the coverage with amorphous carbon on $MoS_2$ surface. The dopants were indeed found to behave as defective sites in the $MoS_2$ lattice, which caused lower adsorption energy and provided anchoring points for accumulating amorphous carbon. The carbon coverage starts in the doped area (Fig. 4a to 4c). By *in situ* EELS monitoring (Fig. S9), we confirmed that the amorphous carbon moved and eventually anchored at the Re dopant sites. The carbon coverage areas are very localized and extend over a few atomic distances from the dopant sites only. This phenomenon directly proves that the dopants cause the alteration of local physical/chemical properties of $MoS_2$ layers.

The C agglomeration process is further corroborated by DFT calculations. We first note that the calculated C adatom diffusion barrier on $MoS_2$ is 0.81 eV, evidencing that C atoms are mobile under the experimental conditions. As illustrated in Figs. 4(d) and (e), the C atom agglomeration around both Re@Mo and Re-Mo is energetically favored. Due to its under-coordination, the Re-Mo adatom binds strongly to the surrounding C atoms even in the case of $C_6$ hexagon. This is in agreement with experiments, where clusters formed more readily around the Re-Mo impurities. Also in the case of Re@Mo, the C adatoms prefer the sites around the Re impurity, but once the C atoms form larger stable clusters, their binding to the impurity becomes fairly weak and is likely dominated by van der Waals forces for larger flakes. In fact, on the basis of these results, we envisage possibility of Re@Mo behaving as a catalytically active site.

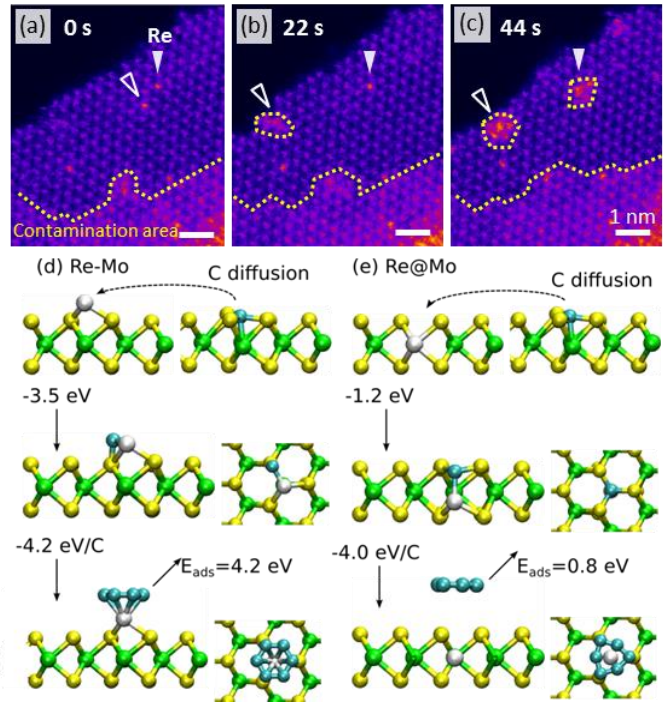

**Figure 4.** Dynamic behavior of carbon contaminations on Re-doped $MoS_2$ surface. (a-c) Three sequential ADF images with Re dopants are indicated by arrows. The yellow dotted regions indicate the area covered by amorphous carbon contaminations. (d-e) Illustration of the C segregation processes around Re-Mo (d) and Re@Mo (e) impurities, and the related energies as obtained from DFT calculations. $E_{ads}$ corresponds to the energy required to remove the molecule from the surface.

In summary, the successful doping of $MoS_2$ with Re (n-type) and Au (p-type) has been demonstrated suggesting the possibility of nano-electronic devices made of doped single-layered $MoS_2$. In comparison with the graphene electronics, the doped $MoS_2$ layers will find broader applications with their suitable bandgaps for nano-electronics. We performed *in situ* experiments providing direct observations of how single dopant atoms interact with other dopants or impurities (such as carbon). Re substitution in Mo site is proved to lower the doping energy for adjacent sites, which results in Re adatom segregation and may explain the difficulties in achieving a high compositional rate of homogeneous Re doping in $MoS_2$. Because the doping sites exhibit local enhancement of chemical reactivity, they likely provide catalytically active sites. Furthermore, the doped $MoS_2$ surface can be used as an ideal specimen support to fix individual molecules for molecular imaging or bio-sensor measurements.[27]

**Experimental Section**

$MoS_2$ single-crystals doped with 0.5~1 at% of Re or Au (purity: 99.99%) were synthesized by the chemical vapor transport (CVT) method with $Br_2$ and $I_2$ as transport agents, respectively.[9,23] The suspended single-layered $MoS_2$ flakes were then prepared by micromechanical cleavage and transferred to a TEM grid (quantifoil) (Fig. S1). ADF imaging was carried out in an aberration-corrected STEM JEOL-2100F equipped with a DELTA corrector and a cold field emission gun operated at 60 kV which is well below the threshold of knock-on damage of $MoS_2$.[28] The scanning rate was set ranging widely from 8 to 128 μs/pixel. The EDX was performed by a JEOL ARM-200F microscope operating at 80kV, spot 4C, 0.5 second/pixels, 50 steps line scan.

In situ experiment of surface coverage of carbon on the $MoS_2$ surface has been performed in TEM. We take advantage of the inevitable carbon deposition in the TEM chamber, the surface coverage rate of carbon can be reduced as slow as possible by carefully chosen experimental conditions. Fresh specimen is required. The single-layered $MoS_2$ flakes transferred from the $SiO_2$ surface to the quantifoil by 2-Propanol and cleaned by Chloroform. Then the specimen were cleaned by baking in air at 200°C for 10 min just before inserting to the TEM chamber after. The JEOL double tilt holder was pre-cleaned by ion-plasma (JEOL ion cleaner, 360V, 10 min) after ethanol treatment. The vacuum level in the TEM chamber is $\sim 1.7 \times 10^{-5}$ pa. 15 min of high current e-beam shower (current density ~107 $pA/cm^2$) was introduced to fix the mobile amorphous carbon and to reduce its migration toward the e-beam scanned area.

Image simulations were carried out by using the MacTempasX software packages. The ADF images were recorded at a convergence angle of $\alpha$ = 24 mrad and an annular detector inner semiangle of $\beta_1$ = 55 mrad. STEM simulation parameters were using a probe size of 1Å, beam current = 15 pA, and microscope focus value = -2 nm with possible specimen inclinations taken into account. For the DFT calculation, the system was modeled using $5 \times 5$ supercell of single-layered $MoS_2$. The Perdew-Burke-Ernzerhof form of the exchange-correction functional[29] was used. Brillouin zone sampled with a $2 \times 2 \times 1$ k-point mesh and 500 eV plane wave cutoff were found to yield converged results.

**Acknowledgements.**

YCL and KS acknowledge a support from JST Research Acceleration programme, and HK and AVK from the Academy of Finland through projects 218545 and 263416. We further thank CSC Finland for generous grants of computer time. Dr. Bernhart Schaffer is gratefully acknowledged for letting us to use his script to record sequential ADF images. DOD and YSH acknowledge the support of the National Science Council of Taiwan under Projects NSC 100-2112-M-011-001-MY3 and NSC 101-2811-M-011-002.

# Supporting Information for

# Properties of Individual Dopant Atoms in Single-Layer MoS$_2$: Atomic Structure, Migration, and Enhanced Reactivity

Yung-Chang Lin,[1] Dumitru O. Dumcenco,[2] Hannu-Pekka Komsa,[3] Yoshiko Niimi,[1] Arkady V. Krasheninnikov,[3] Ying-sheng Huang,[2] Kazu Suenaga[1]*

[1]National Institute of Advanced Industrial Science and Technology (AIST), Tsukuba 305-8565, Japan

[2]Department of Electronic Engineering, National Taiwan University of Science and Technology, Taipei 10607, Taiwan

[3]Department of Physics, University of Heisinki, P.O. Box 43, Fl-00014 Helsinki, Finland

**Contents:**

1. Preparation of large area single-layerd MoS$_2$ specimens.
2. EDX chemical analysis on single dopant atoms
3. STEM simulation and the identification of the atomic position of dopants in MoS$_2$.
4. Migration behaviors of Re and Au adatoms.
5. DFT calculation of formation energy of Re complex dopants.
6. *In situ* experiment for interaction between dopant site and carbon impurity.

1. **Preparation of large area single-layerd MoS₂ specimens.**

Re-doped and Au-doped single crystal $MoS_2$ were transferred to silicon substrate with 300 nm thermal oxide. The number of $MoS_2$ layers was preliminary identified by color contrast in optical microscope. The target flakes were transferred to Mo quantifoil grid [1] and the exact number of layers was confirmed by the Z-contrast of ADF images. Fig. S1a and S1b show the ADF images of large area single- and few-layers of Re-doped and Au-doped $MoS_2$ on quantifoil. Fig. S1c and S1d are examples of ADF images of layer-stacked $MoS_2$ flakes.

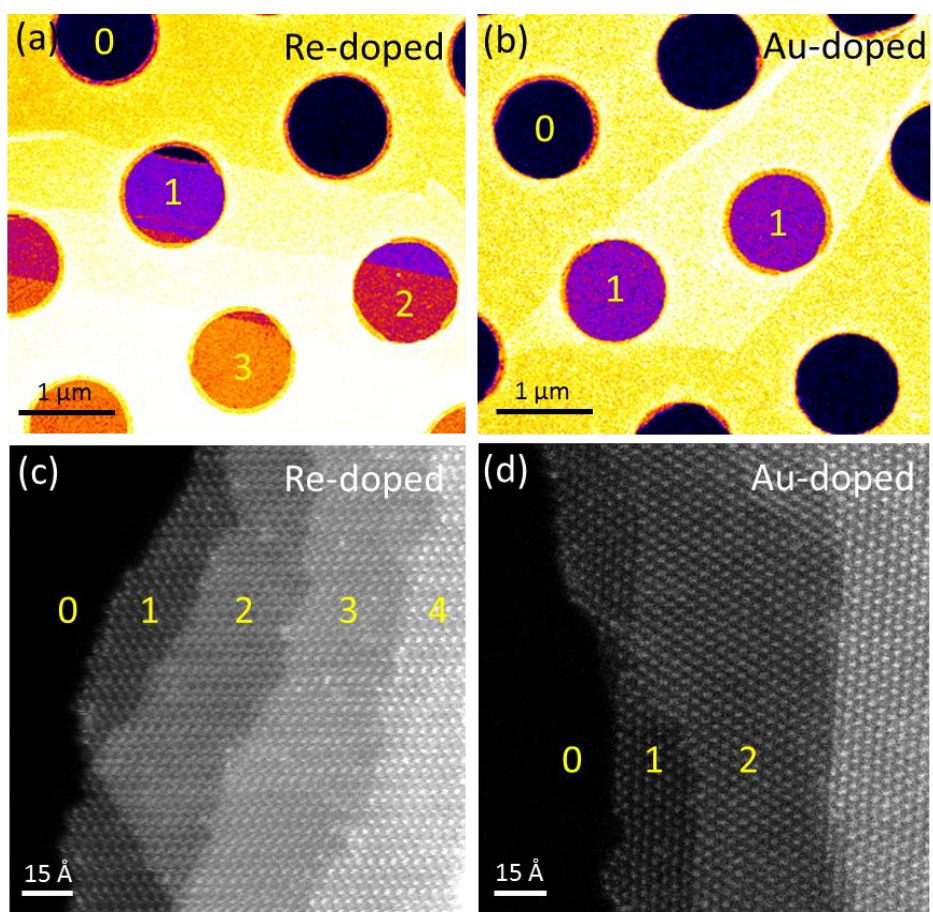

**Figure S1.** (a,b) ADF images of large area single- and few-layers of Re-doped and Au-doped $MoS_2$ transferred onto quantifoil. (c,d) The atomic resolution ADF image of Re-doped and Au-doped few-layer stacked $MoS_2$ flakes. The atomic element arrangements and the number of $MoS_2$ layers can be clearly distinguished by the Z-contrast.

## 2. EDX chemical analysis on single dopant atoms

To confirm the chemistry of examined atoms, the single atom EDX has been carried out [2]. Fig. S2 and S3 show the EDX line scan analysis of single Re and Au atom.

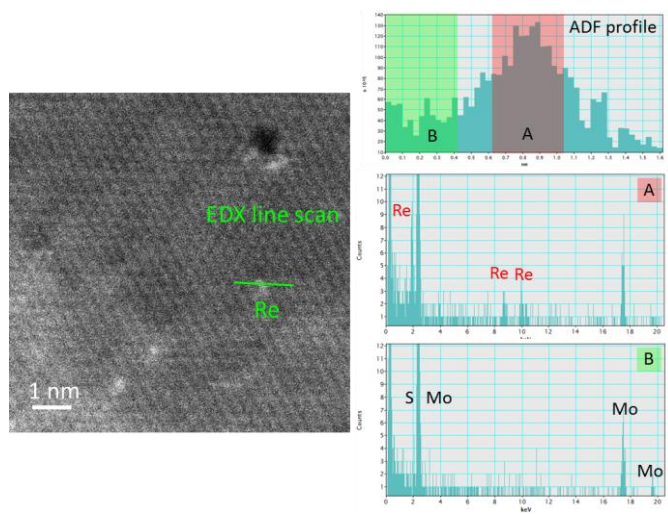

**Figure S2.** EDX chemical analysis of single Re dopant atoms. The e-beam was scanned on the indicated blue green line across a dopant atom (left). The corresponding ADF profile (up), the EDX signal on dopant atom (region A, middle) and off (region B, down) are shown in the right hand side. The EDX spectrum clearly shows the Re signals confirming the dopant element.

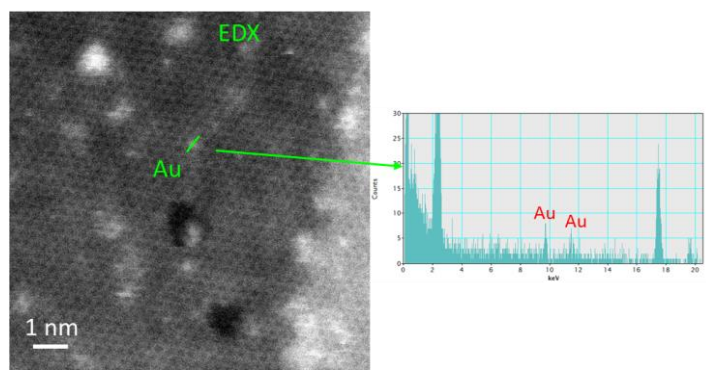

**Figure S3.** EDX on single Au dopant atom. The line-profile is now shown here because the Au atoms tend to move during the line-scan. Nevertheless, the clear signature of Au dopant atom is observed in EDX spectrum (right).

## 3. STEM simulation and the identification of the atomic position of dopants in MoS$_2$.

The simulation ADF intensity is sensitive to the defocus value and the specimen inclination. For Re-doped MoS$_2$, 394 of Re@Mo, 34 of Re-Mo, and 17 of Re-S were observed among 53 individual sequential images. The atomic positions of the Re dopants were confirmed by fitting the experimental ADF profiles with the simulated ones with possible specimen inclinations taken into account [Fig. S4]. Note here that, Re-S configuration can only be observed in the vicinity of Re@Mo [Movie M1], also the Re-S+Re@Mo was also observed [Movie M2].

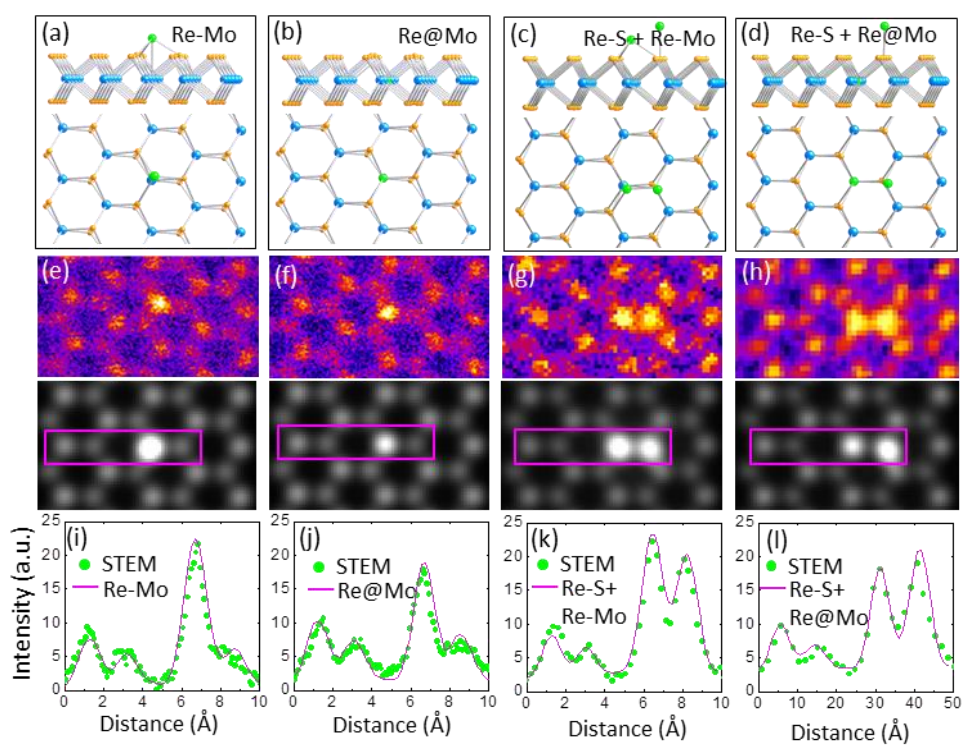

**Figure S4.** Atomic structures of doped MoS$_2$ calculated by DFT for (a) Re adatom on Mo site (Re-Mo), (b) Re substitution at Mo atom (Re@Mo), as well as the complex dopant structure of (c) Re-S+Re-Mo and (d) Re-S+Re@Mo. (e)-(h) The ADF images (the upper panels) and the corresponding simulation images (the lower panels). (i)-(l) The profiles of experimental ADF counts and the fitting of simulation line profiles for each case. A small inclination (4° around x-axis and 4° around y-axis) has been considered to the modeling calculations (a-b); while (c-d) are tilted by 6° around x-axis and 4° around y-axis.

` 19

On the other hand, for the Au-doped single-layered $MoS_2$, 180 of Au-adatoms [Movie M3] and 10 of Au@Mo [Movie M4] were observed among 32 sequential images. Fig. S5 shows the ADF images of distinct atomic positions and the ADF profiles fitting with STEM simulations. The commonly found stacking order of few-layer $MoS_2$ is 2H-stacking [3]. Interestingly, the stacking order of the bilayer region in the Au-doped $MoS_2$ was occasionally found 3R-stacking. The stacking alternation might due to the intercalation of Au adatoms [4]. Fig. S6 shows an ADF image of single-layered Au-doped $MoS_2$ with 3R-stacking bilayer region. Its atomic model and ADF simulation are presented to confirm the 3R-stacking structure.

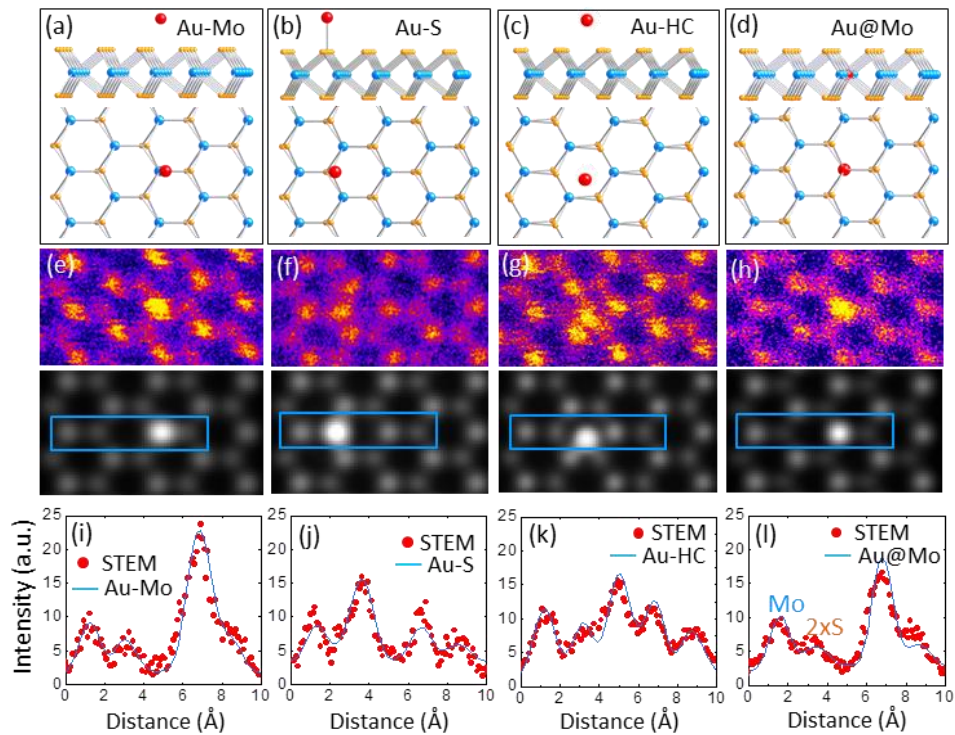

**Figure S5.** (a) Au adatom on Mo site (Au-Mo), (b) Au adatom on S site (Au-S), (c) Au adatom on HC site (Au-HC), and (d) Au substitution at Mo atom (Au@Mo). (e-h) The ADF images (the upper panels) and the corresponding simulation images (the lower panels). (i-l) The profiles of experimental ADF counts and the fitting of simulation line profiles for each case. A small inclination (6° around x-axis) has been applied to the modeling calculations (a), (b) and (d); while (c) are tilted by 1° around x-axis and -6° around y-axis.

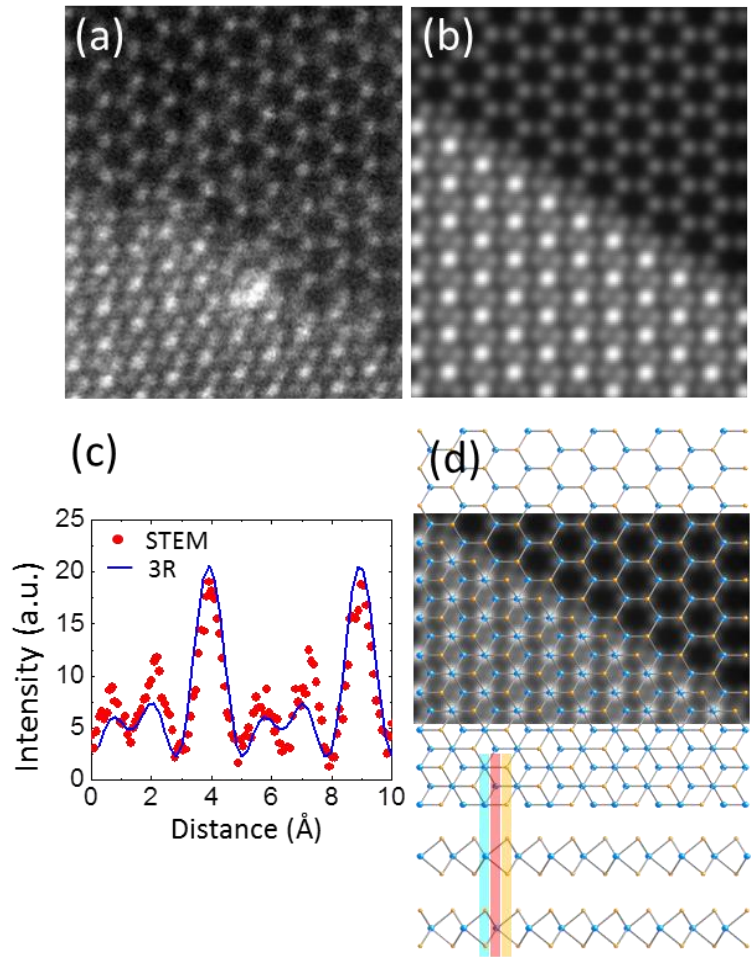

**Figure S6.** (a) The ADF image of clipped from Fig. 2(d). (b) The simulated ADF image. (c) The profile of experimental ADF counts and the fitting of simulation line profile. (d) The atomic model of 3R-stacking bilayer $MoS_2$ superposed to the simulated ADF image. The stacking sequence of 3R-stacking bilayer $MoS_2$ [S-Mo-S] is [A-B-A] and [C-A-C] for the first and second layer, respectively.

## 4. Migration behaviors of Re and Au adatoms.

Under the STEM observations at room temperature, the dopant atoms on MoS$_2$ can become mobile by receiving kinetic energy ($E_k$) from the focused incident electron beam. The energy transferred from the 60 kV electron beam to Re and Au atoms is $E_k$ = 0.75 and 0.71 eV, respectively [5,6]. The binding energy calculated with respect to the energy of the isolated dopant atom is larger than the maximum transferred energy. Such small energies make it unlikely to sputter the dopants from the substitutional or even adatom sites, but can facilitate diffusion. For the case of the Au adatom, the formation energies on the three adatom sites are all very similar and thus migration among them is likely to occur on the MoS$_2$ surface. For Re adatom, there exists a single site with strong binding, which limits the migration and keeps the migration rate small.

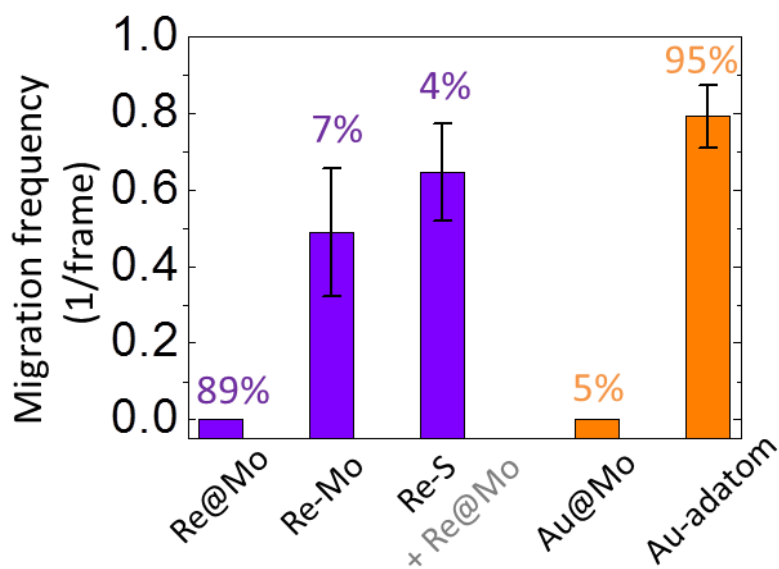

**Figure S7.** The migration frequencies of Re atoms and Au atoms. 89% of Re dopants are found as Re@Mo. Such substitutional dopant atoms do not migrate. On the contrary, 95% of Au dopants are adatoms and tend to migrate frequently.

## 5. DFT calculation of formation energy of Re complex dopants.

Energy differences for the formation of various Re-related complexes are given in Table S1. The calculation results show that all the adatoms (Re-Mo, Re-S, and C-Mo) prefer the sites close to Re@Mo thus forming complex dopant structures.

| Isolated structure ($E_{isolated}$) | Final structure ($E_{final}$) | $\Delta E_f$ (eV) |
|---|---|---|
| [Re-Mo] + [Re@Mo] | [Re-Re@Mo] | -0.96 |
| [Re-Mo] + [Re@Mo](+1) | [Re-Re@Mo](+1) | -0.23 |
| [Re-S] + [Re@Mo] | [Re-S + Re@Mo] | -0.43 |
| [Re-S] + [Re@Mo](+1) | [Re-S + Re@Mo](+1) | -0.30 |
| [Re@S] + [Re@Mo] | [Re@S + Re@Mo] | -0.40 |
| [Re@S] + [Re@Mo](+1) | [Re@S + Re@Mo](+1) | -0.28 |
| [Re-Mo] + [$V_s$ + Re@Mo] | [Re@S + Re@Mo] | -2.18 |
| [Re-Mo] + [$V_s$ + Re@Mo](+1) | [Re@S + Re@Mo](+1) | -2.93 |
| [C-Mo] + [Re@Mo] | [C-Re@Mo] | -1.18 |
| [C-Mo] + [Re@Mo](+1) | [C-Re@Mo](+1) | -0.56 |
| 6[C-Mo] + [Re@Mo] | [$C_6$-Re@Mo] | -21.21 |
| [C-Mo] + [Re-Mo] | [C-Re-Mo] | -3.47 |
| 6[C-Mo] + [Re-Mo] | [$C_6$-Re-Mo] | -24.61 |

**TABLE S1.** Formation energy difference ($\Delta E_f$) between the destined isolated structures ($E_{isolated}$) and the final structures ($E_{final}$), i.e., $\Delta E_f = E_{final} - E_{isolated}$. Note here that, [Re-Mo]+[Re@Mo] stands for an Re adatom on top of a random Mo site in the MoS$_2$ lattice with an Re substitutional dopant at Mo site; [Re-Re@Mo] stands for Re adatom on top of the Re@Mo; [Re-S+Re@Mo] stands for the Re adatom on S site next to a Re@Mo; [$V_s$+Re@Mo] stands for a complex structure of a Sulfur vacancy next to the Re@Mo; and [C-Re@Mo] stands for the carbon adatom on top of the Re@Mo. [$C_6$-Re@Mo] stands for the 6 carbon adatoms on top of the Re@Mo. The calculations covered either charge neutral or +1 positive charged for each equivalent structure. The C atom structures are visualized in Fig. 4.

The energetics of these processes can be understood simply by considering the charge transfer processes (i.e., electronegativities of the constituent species) as quantified here by an analysis of the Bader charges. Generally, Re adatoms are keen to donate electrons to S and C. In the case of Re-S, Re atom has charge +0.7e and the S atom under it -0.3e (as compared to S in pristine $MoS_2$). In the case of Re-Mo, Re atom has charge +1.1e and the three close-by S atoms -0.2e. Carbon atom in C-Mo has charge -0.6e. Atomic charges around neutral Re@Mo are very similar to pristine $MoS_2$ monolayer, except for its immediate neighborhood: Re atom becomes slightly negative (with respect to Mo) and the nearest neighbor S slightly positive. The defect wave function is fairly delocalized in the neighborhood of Re@Mo.

The charge transfer processes during the formation of [Re-S+Re@Mo] complex are plotted in Fig. S8 with respect to isolated Re-S and Re@Mo impurities. In the +1 charge state, the electropositive Re adatom may donate electrons to the S atom below and will consequently have stronger binding to these sites than in the pristine system. In the neutral charge state, there is electron transfer to the complex from its neighborhood, which we assign to the depletion of charge from the shallow Re@Mo defect state.

In the case of impurity segregation, as C adatom moves to Re@Mo site, it obtains charge from the Re atom and binds rather strongly with it. Furthermore, Re adatoms are undercoordinated and thus readily form bonds with other adatoms, which explains the collection of C adatoms around Re-Mo.

We note that the type and role of charge transfer is also evidence by the change in the energy as the charge state of the Re@Mo changes. Adatoms binding directly to Re@Mo rather than to the surrounding S atoms (e.g. C-Mo and Re-Mo) are more sensitive to the charge state of the system.

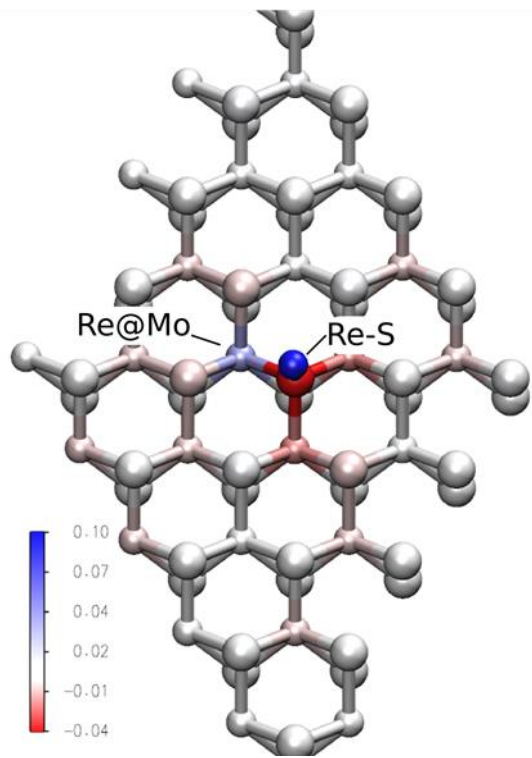 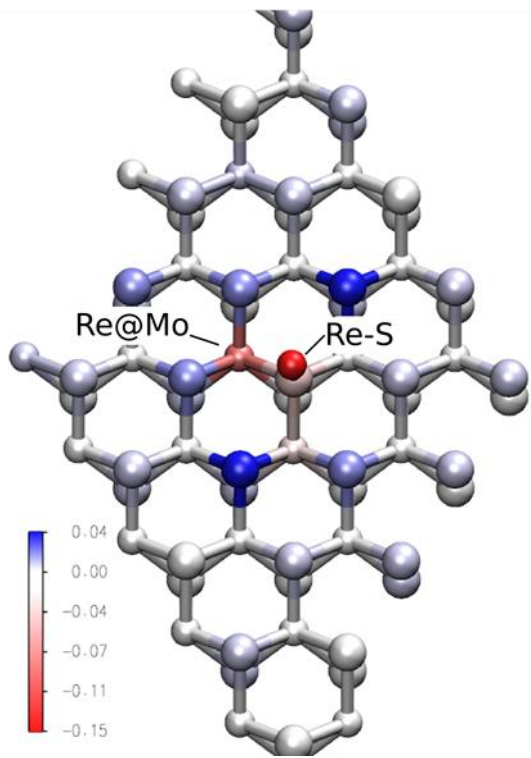

**Figure S8.** Charge transfer during the formation of [Re-S+Re@Mo] complex from its isolated constituents at the (a) +1 and (b) neutral charge states. The charge transfer at each site is calculated as C[Re-S+Re@Mo]+C[bulk]-C[Re-S]-C[Re@Mo], where C correspond to the charge at each atom.

## 6. *In situ* experiment for interaction between dopant site and carbon impurity

Fig. S9 shows time sequence ADF images with the simultaneous EELS acquiring. The carbon contamination aggregated to the vicinity of Re dopants and a noticeable carbon signal detected by EELS (Fig. S9c) in about 20 seconds.

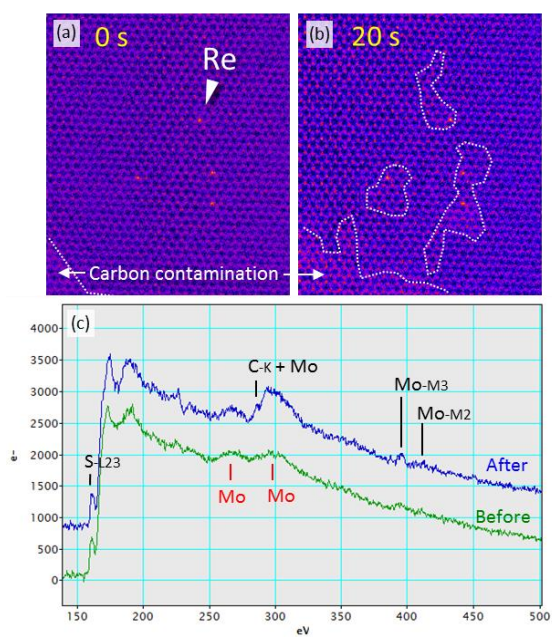

**Figure S9.** (a)(b) Time sequence ADF images of single-layered Re-doped $MoS_2$. The carbon contamination area is surrounded by dotted line. (c) The simultaneously acquired EELS with the sequence ADF images. The increasing of carbon signal at 285 eV is confirmed.